# SHAPE SELECTION IN SYNTHESIS OF COLLOIDS AND NANOPARTICLES

**Igor Sevonkaev**  and  **Vladimir Privman**

Department of Physics, Clarkson University, Potsdam, NY 13699-5721, USA.  E-mail: privman@clarkson.edu

## Introduction

We address a long-standing problem in colloid and nanoparticle science: particle shape selection in solution synthesis [1]. *Our main finding is that a proper balance of two processes, preferential attachment of transported monomers at the protruding features of the growing cluster and monomer rearrangement at the cluster surface, can yield a well-defined particle shape for a large interval of times, and persisting for sizes much larger than the original seed.*

Precipitation in homogeneous solutions has been widely used in preparation of uniform particles because of its experimental versatility [2]. Therefore, an important theoretical challenge has been understanding the mechanisms of diffusional growth of well-defined particles. Specifically, first advances have been reported [3] in understanding particle size selection, i.e., narrow size distribution in polycrystalline colloid synthesis by aggregation of nanosize precursors and in nanoparticle formation by burst nucleation.

Particle shape selection, however, has not been generally understood. The main difficulty has been classifying and modeling the relevant dynamical processes that combine to yield the shape and morphology. Specifically, in properly designed experiments [2], evenly proportioned polycrystalline colloids are obtained with faces corresponding to densely-packed, low-index crystal planes [1].

Several dynamical processes play a role in fast growth by aggregation of diffusionally transported monomers. These include monomer diffusion in solution, their attachment at and detachment from the growing cluster surface, and monomer motion on the surface. The latter process, as well as detachment-reattachment, ultimately leads to the formation of a compact structure of density close to that of the bulk material. In addition, processes that involve more than a single cluster and transport of entities larger than monomer should also be accounted for.

In this study we selected two key dynamical mechanisms: deposition of monomers at exact lattice locations defined by the original seed of cubic/square shape, and their rearrangement on the surface of the growing cluster. The process of diffusional transport is replaced by an artificial mechanism of monomer flux with preferential attachment along directions of the seed corners. This approximation was tested in two and three dimensions (2d and 3d). A more sophisticated model with particle diffu-sion properly treated in a continuum 3d space, and with only the lattice structure imposed by the original seed, has recently been treated in [1]. The main conclusion of both studies has been that cluster growth without development of large defects can yield well-defined particle shapes.

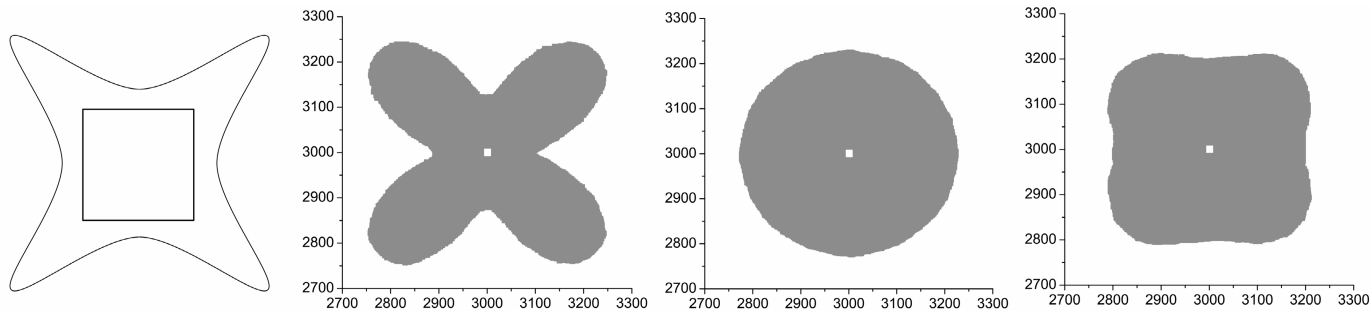

**Figure 1**. Left diagram: GDF with the peaks at the corners of the initial seed. Right plots: Growth of 2d particles for 500,000 iterations each, with different choices of σ and ρ. Note that small white squares at the center of each plot represent the initial seed. From left to right, σ = 0.3 & ρ = 0.3, σ = 0.8 & ρ = 0.3, and σ = 0.5 & ρ = 0.3, respectively.

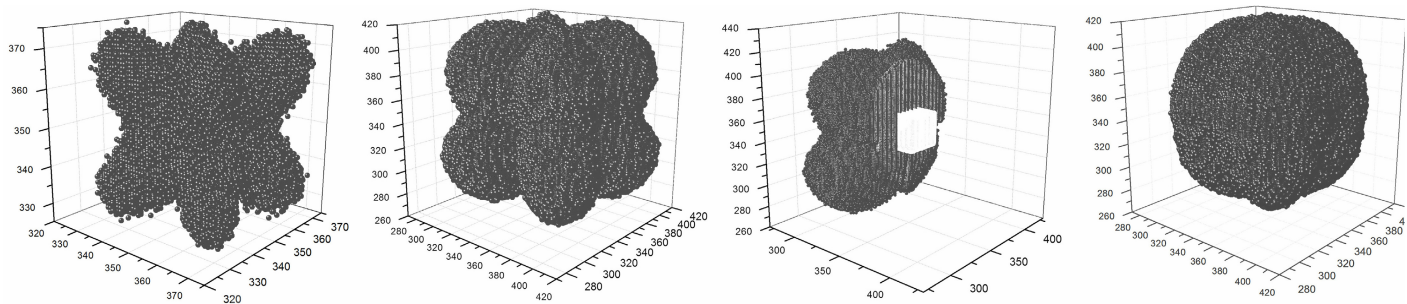

**Figure 2**. Growth of 3d particles for 2,500,000 iterations each, with different choices of σ and ρ. From left to right, σ = 0.2 & ρ = 0.5, σ = 0.3 & ρ = 0.5, again σ = 0.3 & ρ = 0.5 but with part of the cluster removed to expose the original cubic seed, and σ = 0.5 & ρ = 0.5, respectively. Note that the clusters are continuous: The "floaters" (clearly visible in the left panel) are due to the depiction of the cubic-lattice points by spheres which are smaller than the lattice spacing.

**Numerical Approach, Results and Discussion**

Our simulations were carried out by the standard Monte Carlo approach for 2d and 3d systems. The arriving monomers were assumed to deposit irreversibly at the cluster surface according to a Gaussian distribution function (GDF), see Figure 1, simulated by a "polar coordinate" approach [4], with the standard deviation $\sigma$. The rate of rearrangement of monomers at the surface vs. their deposition was controlled by the parameter $\rho$, defined below. Thus, in the present model the effects of the properties of the media and of the monomer and cluster motion were all lumped in the parameters $\sigma$ and $\rho$.

For each choice of $\sigma$ and $\rho$, a uniform random number $0 \leq r \leq 1$ is generated. For $r < \rho$, a new monomer is deposited at the surface of the cluster, at the polar (spherical) direction selected according to the GDF. The motion of this monomer toward the center of the cluster is then carried out along a rectilinear trajectory (with random oscillations about it), as detailed in [5].

For $r > \rho$, a surface monomer, randomly selected from among those earlier deposited, was rearranged (transported) on the cluster surface by a similar process, until it either collided with another monomer or reached a position at the closest possible distance to the center [5].

The model was first tested in 2d, for a $6000 \times 6000$ square lattice. As a typical initial square seed we took a $28 \times 28$ particle, placed at the center of the system. The simulations ran for 500,000 iterations [5]. Three different values of $\sigma$ and $\rho$ (0.3, 0.5, 0.8), i.e., total 9 possible combinations, were tested. The results of 2d modeling with $\rho = 0.3$ are shown in Figure 1 (for $\rho = 0.5$ and 0.8, similar results with minor variations were obtained). As seen in this figure for the case $\sigma = 0.5$ & $\rho = 0.3$, with a proper balance of the two dynamical processes the shape of the initial seed is maintained by the growing cluster up to a large size (as compared to the initial seed), over the simulation time scales.

The two balanced processes were represented by the deposition rate, which favors growth along the directions of the corners of the initial seed, and the rate of the on-surface motion which tends to smooth out the monomer arrangement, driving the cluster shape towards circular. If the processes are not balanced, then one of them "wins" and the cluster either growth protrusions along the preferred directions or becomes circular, as seen in Figure 1.

The 3d simulations were similar, but due to computational resource limitations, we used a $700 \times 700 \times 700$ lattice, again with a $28 \times 28 \times 28$ cubic seed placed at the center. The total number of iterations varied between 2,000,000 and 40,000,000. Our 3d results are illustrated in Figure 2, where the two middle panels show the same cluster grown with properly balanced dynamical processes and as a result approximately maintaining the shape "imprinted" by the original seed. The two other examples in Figure 2 illustrate cluster growth with protrusions or with tendency to become spherical, depending on which process "wins."

In 2d, the core shape was maintained up to sizes approximately two orders of magnitude larger than the initial seed. In 3d, the size ratios tested were much smaller, up to $\times 4$, due to CPU time limitations. Nevertheless, the results clearly indicate that the model is much more sensitive to the choice of the values of $\sigma$ and $\rho$ in 3d than in 2d, which might imply that it could be easier to maintain regimes of well-defined particle growth on 2d substrates than in 3d solutions.

In conclusion, we comment that for general particle growth, the present simplified approach is artificial in the least because the spatial transport is modeled by a distribution controlled by the initial seed rather than by the growing cluster shape. However, as long as the cluster shape follows the original core, we get a glimpse of a possible shape-persistence mechanism over long times and for a range of cluster sizes: the balance of monomer deposition which is sensitive to the shape features and of on-surface monomer rearrangements that favor spherical shapes.

More sophisticated modeling, likely requiring large-scale simulations, is needed to lend credence to this mechanism, which can at best be only approximate because there are other well known modes for cluster shape destabilization. More importantly, further studies are needed to understand the limits of applying such simple rate-balancing expectations and estimate the ranges of time scales and cluster sizes for which shape-persistence can be expected. A step towards a more comprehensive treatment is reported in [1].

We acknowledge funding of our research program by the NSF (grant DMR-0509104).